\documentclass[12pt]{article}
\usepackage{graphicx}
\usepackage{amsmath,amsthm,amsfonts,amssymb}

\def\be{\begin{equation}}
\def\ee{\end{equation}}
\def\mbs{\boldsymbol}
\def\doe{\partial}

\def\dd{{\rm d}} 
\def\df{{\rm d}} 

\begin{document}
\long\def\symbolfootnote[#1]#2{\begingroup\def\thefootnote{\fnsymbol{footnote}}\footnote[#1]{#2}\endgroup}
\def\pamphletheading#1{\paragraph{\Large #1} \hfil \par \noindent}

\def\hsk{\hskip0.1truein}

\title{Analytical Dynamics Development of the Canonical Equations}
\author{John E. Hurtado\thanks{Deputy Director and Chief Technology Officer, Bush Combat Development Complex, Texas A\&M University System; and Professor, Department of Aerospace Engineering, College of Engineering; jehurtado@tamu.edu.} \\ {\it Texas A\&M University, College Station, Texas 77843-3141}}
\date{}
\maketitle

\noindent {\bf Abstract }     It is most common to construct the Hamiltonian function and Hamilton's canonical equations through a Legendre transformation of the Lagrangean function or through the central equation.  These common perspectives, however, seem abstract and detached from classical analytical dynamics.  A new and different approach is presented in which the Hamiltonian function is created as one investigates d'Alembert's equation of motion.  This formulation directly ties the Hamiltonian function and Hamilton's canonical equations to the root of classical analytical dynamics more than any other approach.  

\bigskip


\bigskip 

Like the Lagrangean and the Appellian, the Hamiltonian is yet another descriptive function to generate the equations of motion of a dynamical system.  As mentioned by Pars ([1] p.~184), ``the function contains in itself a complete description of the motions that are possible for the dynamical system.''

It is most common to construct the Hamiltonian function and Hamilton's canonical equations through a Legendre transformation of the Lagrangean function ([2] \S 2.8) or through the central equation ([3] \S 8.2, pp.~1073--1077).  Indeed, Sir Hamilton himself, the eponym of the function and equations, used a version of the central equation approach when he first introduced his ideas ([4] \S 1--3).  Other times, the Hamiltonian function is simply introduced and its origin is not explained at all ([5] \S 7.2, \S 7.5).  Either way, the common perspectives seem abstract and detached from classical analytical dynamics.  

We present a new and different approach in which the Hamiltonian function is created as one investigates d'Alembert's equation of motion.  This formulation directly ties the Hamiltonian function and Hamilton's canonical equations to the root of classical analytical dynamics more than any other approach.  Put another way, this approach brings the canonical equations into the family of classical analytical dynamics.  

This note begins with a brief review of the d'Alembert-Lagrange equation.  Following that is a deliberate change from a traditional motion variable set to a canonical variable set, which sets the stage for discovering the Hamiltonian function and Hamilton's famous canonical equations.  The note ends with a discussion and some concluding remarks.  Einstein’s index summation convention is used throughout; the unfamiliar reader may refer to Frederick and Chang ([6] Ch.~1, passim).

The path to all analytical dynamics forms begins with the d'Alembert-Lagrange equation, also called the fundamental equation, $\left (  m   \ddot{\mbs r} -  \mbs f \right ) \cdot \delta \mbs r =0 $.  (For the simplicity of the current presentation, we consider the motion of a single particle.) Here, $m$ is the particle's mass, $\mbs r$ is the particle's position vector for an inertial observer, and  $\ddot {\mbs r}$, the second inertial time derivative of $\mbs r$, is the particle's inertial acceleration vector.  The vector $\mbs f$ represents the total of all externally applied forces and $\delta \mbs r$ represents the virtual displacement vector.  

The position vector is generally a function of time and a set of coordinates, $\mbs r = \mbs r(t,q)$.  Here, $q = \{q_1, \ldots, q_n\}$ is a general set of unconstrained coordinates where $n$ is the minimum number of coordinates that are needed to completely specify the body's configuration.  The set $\dot q = \{\dot q_1, \ldots, \dot q_n\}$, are the corresponding time derivatives and are commonly called generalized velocities. 

As mentioned by Papastavridis ([3] p.~280), the string of kinematic identities ${\doe \ddot{\mbs r}}/{\doe \ddot q_k} = {\doe \dot{\mbs r}}/{\doe \dot q_k}  = {\doe \mbs r}/{\doe q_k}$ holds true for any well-behaved function of time and generalized coordinates, like the position vector function $\mbs r = \mbs r(t,q)$.  Moreover, the $k$th holonomic Euler-Lagrange operator equals zero when applied to the time derivatives of such functions, as in $E_k(\dot{\mbs r}) \equiv \dd/\dd t \left ( \doe \dot{\mbs r} / \doe \dot q_k \right) - \doe \dot {\mbs r} / \doe q_k=0$.  These integrability conditions are crucial to analytical dynamics.     

Following Likins ([7] p.~181) and Kane ([8] p.~53), the virtual displacement vector $\delta \mbs r$ is an imagined displacement away from the particle's true position in terms of increments in generalized coordinates and is valid at any instant of time, $\delta \mbs r = \left ({\doe \mbs r}/{\doe q_k}\right) \delta q_k$.  Equally, because of the kinematic identities, $\delta \mbs r = \left ({\doe \dot{\mbs r}}/{\doe \dot q_k} \right ) \delta q_k$.  

Either of these expressions for the virtual displacement vector can be used to define the generalized forces, for example $Q_k \equiv \mbs f \cdot \doe \dot{\mbs r} / \doe \dot q_k$, and this in turn leads to a restatement of the d'Alembert-Lagrange equation, $\left (  m   \ddot{\mbs r} \cdot \doe \dot{\mbs r} / \doe \dot q_k -  Q_k\right ) \delta q_k = 0$.  And because each generalized coordinate increment $\delta q_k$ is independent from the others and arbitrary in amount, it must be that the parenthetical quantity is identically zero for each value of $k$. 
\be
m   \ddot{\mbs r} \cdot \frac{\doe \dot{\mbs r}}{\doe \dot q_k} = Q_k  
\ee

At the heart of analytical dynamics is the realization that $m \ddot {\mbs r} \cdot {\doe \dot{\mbs r}}/{\doe \dot q_k}$ can be transformed into calculations on a descriptive function.  For Lagrange, that function is the kinetic energy function whereas for Appell, that function is the Appellian.  For Hamilton, that function is the Hamiltonian and the journey from $m \ddot {\mbs r} \cdot {\doe \dot{\mbs r}}/{\doe \dot q_k}$ to the discovery of, and calculations on this function is precisely the remaining focus of this note.  

Consider the calculation of $m \ddot{\mbs r} \cdot {\doe \dot{\mbs r}}/{\doe \dot q_k}$.  
\begin{align}
m \ddot{\mbs r} \cdot \frac{\doe \dot{\mbs r}}{\doe \dot q_k}  & = \frac{\df}{\df t} \left ( m \dot{\mbs r} \cdot \frac{\doe \dot{\mbs r}}{\doe \dot q_k} \right ) - m \dot{\mbs r} \cdot \frac{\df}{\df t} \left ( \frac{\doe \dot{\mbs r}}{\doe \dot q_k} \right ) \nonumber \\
\noalign{\medskip}
& = \frac{\df}{\df t} \left ( m \dot{\mbs r} \cdot \frac{\doe \dot{\mbs r}}{\doe \dot q_k} \right ) - m \dot{\mbs r} \cdot \frac{\doe \dot{\mbs r}}{\doe q_k}
\end{align}
We may rewrite this expression by using the kinetic energy function, $T(t,q,\dot q)$ $= \frac{1}{2} m \dot{\mbs r} \cdot \dot{\mbs r}$, and by introducing the concept of {\it conjugate momenta}, which here are defined as the projections of the momentum vector $m \dot {\mbs r}$ along the partial velocity vector directions, $p_k (t,q,\dot q)\equiv m \dot{\mbs r} \cdot \doe \dot{\mbs r} / \doe \dot q_k$.
\be
m \ddot{\mbs r} \cdot \frac{\doe \dot{\mbs r}}{\doe \dot q_k}= \dot p_k - \frac{\doe T}{\doe q_k}
\ee

Hamilton's equations are ordinary differential equations in the canonical variable set $(q, p)$ rather than the traditional variable set $(q, \dot q)$.  Thus, in moving ahead we must establish the explicit relationship between the generalized velocities and conjugate momenta, and then use this relationship to transform Eq.~(3).

To begin, note that the velocity vector may be written in terms of generalized coordinates, generalized velocities and partial derivative vectors, $\dot{\mbs r} = \left ({\doe \dot{\mbs r}}/{\doe \dot q_i} \right ) \dot q_i + {\doe \mbs r}/{\doe t}$.  Using this relationship in the definition of conjugate momenta gives the following:
\be
p_k(t,q,\dot q) = m \dot{\mbs r} \cdot \frac{\doe \dot{\mbs r}}{\doe \dot q_k} = m \frac{\doe \dot{\mbs r}}{\doe \dot q_i} \, \dot q_i \cdot \frac{\doe \dot{\mbs r}}{\doe \dot q_k}  + m \frac{\doe \mbs r}{\doe t} \cdot \frac{\doe \dot{\mbs r}}{\doe \dot q_k} = M_{ki} \dot q_i + G_k 
\ee
The definitions of $M_{ik} = M_{ki} = M_{ki}(t,q)$ and $G_k = G_k (t,q)$ are clear.  The inverse of this relationship is straightforward, where $A_{jk}$ is the inverse of $M_{ki}$.  
\be
\dot q_j(t,q,p) = A_{jk} \left ( p_k - G_k \right ) 
\ee

Having the relationships between the generalized velocities and conjugate momenta, we now consider the kinetic energy function in terms of the different variable sets.  The kinetic energy function is classically defined as $T = \frac{1}{2} m \dot{\mbs r} \cdot \dot{\mbs r}$.  Note that if $\dot{\mbs r} = \dot{\mbs r} (t,q,\dot q)$, then $T = T(t,q,\dot q)$; but if $\dot{\mbs r} = \dot{\mbs r} (t,q,\dot q (t,q,p)) = \dot{\mbs r}(t,q,p)$, then $T = T(t,q,p)$.  For clarity we will begin using a superscript $\star$ notation on a scalar or vector function when the function depends on the canonical variables $(q,p)$.  

The explicit forms of the kinetic energy can be determined, and we begin with the form in terms of traditional variables.   
\begin{equation}
T (t,q,\dot q) = \frac{1}{2} m \dot{\mbs r} \cdot \dot{\mbs r} = \frac{1}{2} M_{ij} \, \dot q_i \dot q_j + G_i \, \dot q_i + T_0 \quad \text{where} \quad  T_0 = \frac{1}{2} m \frac{\doe \mbs r}{\doe t} \cdot \frac{\doe \mbs r}{\doe t}
\end{equation}
Using the relationship between the generalized velocities and conjugate momenta, it is straightforward to determine the kinetic energy function in terms of canonical variables.  
\begin{equation}
T^\star (t,q,p)  = \frac{1}{2} m \dot{\mbs r}^\star \cdot \dot{\mbs r}^\star = \frac{1}{2} A_{ij} \, p_i p_j - \frac{1}{2} G_i A_{ij} G_j + T_0 
\end{equation}

Having these two forms for the kinetic energy function, we return to Eq.~(3), where the kinetic energy function depends on traditional variables, $T = T(t,q, \dot q)$, yet Hamilton's equations are tailored to make use of canonical variables $(q, p)$.  Thus, we use the chain rule of calculus applied to $T^\star (t,q,p)$ to achieve the transition.  
\begin{align}
m \ddot{\mbs r} \cdot \frac{\doe \dot{\mbs r}}{\doe \dot q_k} = \dot p_k - \frac{\doe T}{\doe q_k} = \dot p_k - \frac{\doe T^\star}{\doe q_k} - \frac{\doe T^\star}{\doe p_i} \frac{\doe p_i}{\doe q_k}
\end{align}
We can perform the explicit computations in the far right expression of Eq.~(8) to complete the transition to the canonical variable set $(q,p)$.  In the end, Eq.~(1) becomes the following.
\begin{equation}
\dot p_k = - \frac{1}{2} \frac{\doe A_{ij}}{\doe q_k} p_i p_j + p_i \frac{\doe}{\doe q_k} \left ( A_{ij} G_j \right ) - \frac{\doe}{\doe q_k} \left ( \frac{1}{2} G_i A_{ij} G_j - T_0 \right ) + Q_k 
\end{equation}
Equations (5) and (9) are first order differential equations in canonical variables $(q,p)$ that governing the system's motion.  But they are far from being recognized as Hamilton's canonical equations.  

Although we've successfully transitioned the equations of motion from traditional variables $(q, \dot q)$ to canonical variables $(q,p)$, there are a few remaining steps to arrive at Hamilton's canonical equations.  To focus our presentation, we consider the governing equations of motion in canonical variables $(q,p)$ for the special case of no generalized forces, i.e., $Q_k=0$.  
\begin{align}
& \dot q_k = A_{kj} \left ( p_j - G_j \right ) \nonumber \\
& \dot p_k = \frac{\doe T^\star}{\doe q_k} + \frac{\doe T^\star}{\doe p_i} \frac{\doe p_i}{\doe q_k} = - \frac{1}{2} \frac{\doe A_{ij}}{\doe q_k} p_i p_j + p_i \frac{\doe \left ( A_{ij} G_j \right )}{\doe q_k} - \frac{1}{2} \frac{\doe \left ( G_i A_{ij} G_j \right )}{\doe q_k} + \frac{\doe T_0}{\doe q_k} \nonumber
\end{align}

Hamilton's equations involve computations on a single descriptive function called the Hamiltonian function.  In our pursuit of this new function, we seek an auxiliary scalar function $S^\star(t,q,p)$ whose addition to $T^\star$ becomes a feasible candidate.  This leads us to consider the following.  
\begin{align}
& \dot q_k = A_{kj} \left ( p_j - G_j \right ) = \frac{\doe}{\doe p_k} \left ( T^\star + S^\star \right ) \\
& \dot p_k = \frac{\doe T^\star}{\doe q_k} + \frac{\doe T^\star}{\doe p_i} \frac{\doe p_i}{\doe q_k} = - \frac{\doe}{\doe q_k} \left ( T^\star + S^\star \right )
\end{align}

First consider Eq.~(10) while recalling that $T^\star = \frac{1}{2} A_{ij} \, p_i p_j - \frac{1}{2} G_i A_{ij} G_j + T_0$.  
\begin{align}
& \frac{\doe S^\star}{\doe p_k} = A_{kj} \left ( p_j - G_j \right ) - \frac{\doe T^\star}{\doe p_k} = A_{kj} \left ( p_j - G_j \right ) - A_{kj} p_j = -A_{kj}G_j  \nonumber \\
& \qquad \to \qquad S^\star (t,q,p) = -p_k A_{kj} G_j + s_q(t,q)
\end{align}
Here, $s_q(t,q)$ is a new function, independent of conjugate momenta, that arises from integration.  

Next consider Eq.~(11).  Using the computations we have performed so far leads to another expression for $S^\star$.  
\begin{align}
& \frac{\doe S^\star}{\doe q_k} = -2 \frac{\doe T^\star}{\doe q_k} - \frac{\doe T^\star}{\doe p_i} \frac{\doe p_i}{\doe q_k} = -p_i \frac{\doe \left ( A_{ij} G_j \right ) }{\doe q_k} + \frac{\doe \left ( G_i A_{ij} G_j - 2T_0 \right ) }{\doe q_k} \nonumber \\
& \qquad \to \qquad S^\star (t,q,p) = -p_i A_{ij} G_j + G_i A_{ij} G_j - 2T_0 + s_p(t,p)
\end{align}
Like $s_q$, the function $s_p(t,p)$ is a new function, independent of generalized coordinates, that arises from integration.  

Comparing Eqs.~(12) and (13) reveals $s_q(t,q) = G_i A_{ij} G_j - 2T_0$ and $s_p(t,p) = 0$, and therefore $S^\star (t,q,p) = -p_i A_{ij} G_j + G_i A_{ij} G_j - 2T_0$.  We now have our first glimpse of the Hamiltonian function.  
\be
H^\star (t,q,p) \equiv T^\star + S^\star = \frac{1}{2} A_{ij} \, p_i p_j - p_i A_{ij} G_j + \frac{1}{2} G_i A_{ij} G_j - T_0
\ee

Equation (14) is a Hamiltonian function $H^\star(t,q,p)$, which now allows us to write one form of Hamilton's equations wherein we reconsider the presence of generalized forces. 
\begin{equation}
\begin{array}{l}
\dot q_k = A_{kj} \left ( p_j - G_j \right )  \\
\noalign{\medskip}
\dot p_k = \displaystyle \frac{\doe T^\star}{\doe q_k} + \frac{\doe T^\star}{\doe p_i} \frac{\doe p_i}{\doe q_k} + Q_k
\end{array}
\qquad \to \qquad
\begin{array}{l}
\dot q_k = \displaystyle \frac{\doe H^\star}{\doe p_k}  \\
\noalign{\medskip}
\dot p_k = \displaystyle - \frac{\doe H^\star}{\doe q_k} + Q_k
\end{array}
\end{equation}

As mentioned earlier, the classical definition of the kinetic energy function is the coordinate-free expression $T = \frac{1}{2} m \dot{\mbs r} \cdot \dot{\mbs r}$, which holds true regardless of the variable set.  We now show that the auxiliary function $S^\star$ can be written in a similar coordinate-free manner. 
  
Consider the auxiliary function $S^\star$ while folding in the relationship between the generalized velocities and conjugate momenta.  
\begin{align}
S^\star (t,q,p) & = -p_i A_{ij} G_j + G_i A_{ij} G_j - 2T_0 = -G_j A_{ji} \left ( p_i - G_i \right ) - 2 T_0 \nonumber \cr
& = -G_j \dot q_j - 2T_0 = - m \frac{\doe \dot{\mbs r}}{\doe \dot q_j} \cdot \frac{\doe \mbs r}{\doe t} \, \dot q_j - m \frac{\doe \mbs r}{\doe t} \cdot \frac{\doe \mbs r}{\doe t} \cr 
& = - m \left ( \frac{\doe \dot{\mbs r}}{\doe \dot q_j} \, \dot q_j + \frac{\doe \mbs r}{\doe t} \right ) \cdot \frac{\doe \mbs r}{\doe t} \cr
\to \quad S & = -m \dot{\mbs r} \cdot \frac{\doe \mbs r}{\doe t} 
\end{align}

Having the kinetic energy function $T$ and auxiliary function $S$ in coordi\-nate-free forms allows us to write the Hamiltonian function $H$ in a beautiful and concise similar way.  
\be 
H = \frac{1}{2} m \dot{\mbs r} \cdot \dot{\mbs r} - m \dot{\mbs r} \cdot \frac{\doe \mbs r}{\doe t} \quad \text{or} \quad H = T - p_o
\ee
Here, we're suggesting that because $m \dot{\mbs r} \cdot {\doe \dot{\mbs r}}/{\doe \dot q_k}$ defines $p_k$, then $m \dot{\mbs r} \cdot {\doe \mbs r}/{\doe t}$ could define $p_o$ because of the striking similarity in appearance.  Furthermore, because $p_o \equiv m \dot{\mbs r} \cdot {\doe \mbs r}/{\doe t}$ stems from the auxiliary function $S$ and has the units of energy, it should be called {\it the auxiliary energy}. 

d'Alembert's equation of motion has led to the current form of Hamilton's equations.  (See the far right set of equations in Eq.~(15) and Eq.~(16).)  When the generalized forces $Q_k$ are all derivable from a (collection of) potential energy function(s), $V = V(t,q)$, so that $Q_k = - \doe V / \doe q_k$, one can reassign (or re-set) the definition of the Hamiltonian function, $H \leftarrow H + V$, and write {\it Hamilton's famous canonical equations}.
\be
\dot p_k = - \frac{\doe H^\star}{\doe q_k} \qquad ; \qquad \dot q_k = \frac{\doe H^\star}{\doe p_k}
\ee
\vskip-.2truein
\be
\text{where} \quad {H} = E - p_o = \frac{1}{2} m \dot{\mbs r} \cdot \dot{\mbs r} + V - m \dot{\mbs r} \cdot \frac{\doe \mbs r}{\doe t}
\ee
Here, $E=T+V$ is the system energy and we are following Meirovitch ([9] \S2.13) who reserves the name canonical for this case.  

Hamilton's famous canonical equations cover the case wherein {\it any and all\/} generalized forces are derivable from potential energy functions.  If only {\it some\/} generalized forces are derivable from potential functions, then $H$ accounts for the potential energy contributions like before and the remaining nonpotential generalized forces are simply added to the right side of the kinetic equation.  It is customary to drop the canonical name in that case. 

The kinetic energy function, the Appell function, and the Hamiltonian function are each descriptive functions that naturally appear as one investigates d'Alembert's equation of motion.  
\begin{align*}
\begin{array}{ll}
\text{Lagrange:} & \displaystyle m \ddot{\mbs r} \cdot \frac{\doe \dot{\mbs r}}{\doe \dot q_k} = \frac{\df}{\df t} \left ( \frac{\doe T}{\doe \dot q_k} \right ) - \frac{\doe T}{\doe q_k} \quad \text{where} \quad T = \frac{1}{2} m \dot{\mbs r} \cdot \dot{\mbs r}  \\
\noalign{\bigskip}
\text{Appell:} & \displaystyle m \ddot{\mbs r} \cdot \frac{\doe \ddot{\mbs r}}{\doe \ddot q_k} = \frac{\doe \cal A}{\doe \ddot q_k} \quad \text{where} \quad {\cal A} = \frac{1}{2} m \ddot{\mbs r} \cdot \ddot{\mbs r} \\
\noalign{\bigskip}
\text{Hamilton:} & \displaystyle m \ddot{\mbs r} \cdot \frac{\doe \dot{\mbs r}}{\doe \dot q_k} = \dot p_k = - \frac{\doe H^\star}{\doe q_k} \quad \text{with} \quad \dot q_k = \frac{\doe H^\star}{\doe p_k} \\
& \hskip1.2truein \text{where} \displaystyle \quad {H} = \frac{1}{2} m \dot{\mbs r} \cdot \dot{\mbs r} - m \dot{\mbs r} \cdot \frac{\doe \mbs r}{\doe t} \\
\end{array}
\end{align*}

Admittedly, the descriptive functions $T$ and $\cal A$ seem to magically appear in their respective developments whereas more effort is needed to coax the appearance of $H^\star$.  Nevertheless, this new approach directly ties the construction of the Hamiltonian function and Hamilton's equations to d'Alembert's equation and consequently to classical analytical dynamics.  

Seeing the Lagrange, Appell, and Hamilton equations side by side, we note that the Lagrangean and Appellian approaches shown here lead to second order differential equations whereas the Hamiltonian approach always and naturally leads to a set of first order differential equations.  We have seen, however, that the Lagrangean and Appellian approaches can also lead to first order differential equations when nonholonomic velocity variables ([3] p.~418-419), also called quasi-velocities or generalized speeds ([5] \S 5.4.5; [10] \S 7.6), are introduced to replace the generalized velocities.  This indicates that the conjugate momenta are a special set of generalized speeds such that the kinetics and kinematics are captured by a descriptive function.  

For the first time, the most famous descriptive functions may be viewed as standing on equal footing.  Additionally, there is likely no other function to be discovered, there is nothing to add to the list.  The reason is because the Lagrangean function is the appropriate descriptive function, to within the addition of a function $\dd f(t,q) / \dd t$, in terms of generalized coordinates and generalized velocities, $L=L(t,q,\dot q)$; likewise, the Appellian function is the appropriate descriptive function, again to within the addition of a function $\dd f(t,q) / \dd t$, in terms of generalized coordinates, generalized velocities, and generalized accelerations,  ${\cal A}={\cal A}(t,q,\dot q,\ddot q)$; and the Hamiltonian function is the appropriate descriptive function in terms of generalized coordinates and other speeds, $H^\star=H^\star(t,q,p)$.  

Interestingly, unlike the Lagrangean and Appellian functions, the Hamiltonian function will not tolerate the addition of a function $\dd f(t,q) / \dd t$: the Hamiltonian is unique in that way.  

Finally, although the development throughout has been for a single particle, it should be clear that the method extends in a straightforward way to a collection of particles, a collection of rigid bodies, or a mix of the two.  


\section*{Funding }
This research did not receive any specific grant from funding agencies in the public, commercial, or not-for-profit sectors.  

\section*{References}
\begin{enumerate}

\item [{[1]}] Pars, L.A., {\it A Treatise on Analytical Dynamics}, Ox Bow Press, Woodbridge, CT, 1981.  

\item [{[2]}] McCauley, J.L., {\it Classical Mechanics: transformations, flows, integrable, and chaotic dynamics}, Cambridge University Press, New York, NY, 1997.  

\item [{[3]}] Papastavridis, J.G., {\it Analytical Mechanics}, Oxford Univ.~Press, New York, NY, 2002.  

\item [{[4]}] Hamilton, W.R., ``Second Essay on a General Method in Dynamics,'' {\it Philosophical Transactions of the Royal Society}, Vol.~125, pp.~95-144.

\item [{[5]}] Schaub, H., and Junkins, J.L., {\it Analytical Mechanics of Space Systems}, AIAA, Reston, VA, 2003.  

\item [{[6]}] Frederick, D., and Chang, T. S., {\it Continuum Mechanics}, Scientific Publishers, Cambridge, MA, 1965.  

\item [{[7]}] Likins, P.W., {\it Elements of Engineering Mechanics}, McGraw-Hill, New York, NY, 1988.   


\item [{[8]}] Kane, T. R., {\it Dynamics}, Holt, Rinehart and Winston, New York, NY, 1968.


\item [{[9]}] Meirovitch, L., {\it Methods of Analytical Dynamics}, McGraw–Hill, New York, NY, 1970.  

\item [{[10]}] Baruh, H., {\it Analytical Dynamics,} McGraw–Hill, New York, NY, 1999.  

\end{enumerate}
\end{document}